\documentclass[twocolumn,usenatbib,usegraphicx ]{mn2e}

\def\Mpc{{\rm Mpc}}
\def\lcdm{$\Lambda {\rm CDM} \,\,$} 
\def\mnras{MNRAS}

\def\aj{AJ}
\def\apj{ApJ}
\def\apjl{ApJL}
\def\apjs{ApJS}
\def\aap{A\&A}
\def\araa{Annual Review of Astronomy and Astrophysics}

\def\pasp{Publications of the Astronomical Society of the Pacific}

\usepackage{epsfig}

\begin{document}
\title{Exploring star formation using  the filaments in the Sloan
  Digital Sky Survey Data Release Five (SDSS DR5)}  
\author[B. Pandey and S. Bharadwaj] {Biswajit Pandey$^1$\thanks{Email:
biswa@iucaa.ernet.in} and Somnath Bharadwaj$^2$\thanks{Email:
somnath@cts.iitkgp.ernet.in} \\
${}^1$ Inter-University Centre for Astronomy and
Astrophysics, Pune 411 007,  India \\${}^2$ Department of Physics and
Meteorology \& Centre for Theoretical Studies, IIT Kharagpur, 
 721 302 , India } \maketitle

\begin{abstract}
We have quantified the average filamentarity of the galaxy
distribution in seven nearly two dimensional strips from the SDSS DR5
using a volume limited sample in the absolute magnitude range $-21 \le
M_r \le -20$. The average filamentarity of star forming (SF) galaxies,
which are predominantly blue, is found to be more than that of other
galaxies which are predominantly red.  This difference is possibly
 an outcome of the fact that blue galaxies have a more filamentary
distribution. Comparing the SF galaxies with only the blue other
galaxies, we find that the two show nearly equal
filamentarity. Separately analyzing the galaxies with high star
formation rates (SFR) and low SFR, we find that the latter has a more
filamentary distribution.  We interpret this in terms of two 
effects (1.)  A correlation between the SFR and
individual galaxy properties  like luminosity  with the high SFR galaxies
being more luminous (2.) A  relation between
the SFR and environmental effects like the  density with the high SFR
galaxies preferentially occurring in high density regions.  These two
effects are possibly not independent and are operating simultaneously.  
We do not find any difference in the filamentarity of SF galaxies and
AGNs. 
\end{abstract}
\begin{keywords}
methods: numerical - galaxies: statistics - 
cosmology: theory - cosmology: large scale structure of universe 
\end{keywords}

\section*{Introduction}
Determining the factors that govern star formation in a galaxy is an
important issue which is expected to shed light on our understanding
of how galaxies are formed. It is believed that there are several
interlinked factors like turbulence, magnetic fields, cosmic rays,
steller winds, supernova explosions, etc. functioning in a galaxy's
interstellar medium (ISM) which together regulate star formation in
the galaxy (see \citealt{larson}, \citealt{mckee} for recent
reviews). There is now mounting evidence that in addition to these
factors operating inside a galaxy, its star formation is also
influenced by external factors. It is quite clear that interactions
between galaxies (e.g. \citealt{byrd}, \citealt{moore}) and the
interactions of a galaxy with the external ambient medium
(e.g. \citealt{gunn}, \citealt{zabludoff}, \citealt{porter}) can
trigger star formation.  There is also observational evidence that
star formation is suppressed in regions where the galaxy density is
high \citep{lewis,gomez,balogh}. This is possibly related to the fact
that other galaxy properties like luminosity (e.g. \citealt{einas2},
\citealt{einas3}, \citealt{park2}), colour (e.g. \citealt{hog1},
\citealt{blan1}) and morphology (e.g. \citealt{dress}, \citealt{goto})
also exhibit environmental dependence.

The analysis of filamentary patterns in the galaxy distribution has a
long history dating back to a few papers in the late-seventies and
mid-eighties by \citet{joe}, \citet{einas4}, \citet{zel},
\citet{shand1} and \citet{einas1}.  Filaments are the most striking
visible patterns in the galaxy distribution (e.g. \citealt{gel},
\citealt{shect}, \citealt{shand2}, \citealt{bharad1}, \citealt{mul},
\citealt{basil}, \citealt{doro2}, \citealt{pimb}). The filamentarity is
found to be statistically significant up to length-scales $80 \,
h^{-1} {\rm Mpc}$ and not beyond \citep{bharad2,pandey}. Our earlier
work (\citealt{pandey1}, hereafter Paper A) shows the degree of
filamentarity to depend on physical properties like the luminosity,
colour and morphology of the galaxies.  In the present work we
investigate the relation between the filamentarity observed in the
galaxy distribution and ongoing star formation activity in the
galaxies.

 A brief outline of our paper follows. Section 2 describes the data
 and  method of analysis, our results  and
 conclusions are presented in Section 3.
We have used a \lcdm cosmological model with
$\Omega_{m0}=0.3$, $\Omega_{\Lambda0}=0.7$ and $h=1$ throughout.

\section*{Data and Method of analysis}

\begin{figure}
\rotatebox{-90}{\scalebox{.34}{\includegraphics{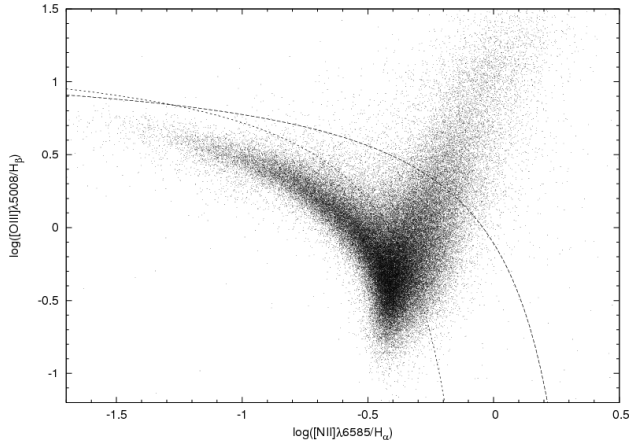}}}
\caption{ BPT diagnostic diagram showing the ratios
$[OIII]/H_{\beta}$ and  $[NII]/H_{\alpha}$ for all
galaxies with $I_\lambda/\sigma_{I_{\lambda}}>2$. The 
theoretical curve separating SF and AGNs (upper-dashed,
\citealt{kewley}) and  the empirical curve (lower-dotted,
\citealt{kauff})  are shown.}
\label{fig:1}
\end{figure}

The Sloan Digital Sky Survey (SDSS,  \citealt{york}) is a
five-passband (u,g,r,i,z) imaging and spectroscopic survey of the
Northern Galactic hemisphere to a limiting Petrosian r band magnitude
$r<17.77$.
Our analysis is limited to  seven   thin strips on the sky drawn  
from the
SDSS DR5 \citep{adel},   
each spanning $90^\circ$ in $\lambda$ and $2^\circ$ in
$\eta$ .
Here $\lambda$ and $\eta$ are survey co-ordinates defined in
\citet{stout}. These strips are identical in sky-coverage as the ones
used in Paper A and are shown in Figure 1 of that paper. Only
galaxies with extinction corrected Petrosian r band magnitude in the
range $ 14.5 \leq m_r \leq 17.77 $ were used. Volume limited samples
are constructed in the same way as discussed in Paper A. The
samples cover r-band absolute magnitude range $-21\leq M_r \leq -20$
and redshift range $0.043657 \leq z \leq 0.114635$ which correspond to
$130 \, h^{-1} \Mpc$ to $335 \, h^{-1} \Mpc$ comoving in the radial
direction. Finally we have 17225 galaxies distributed in seven 
strips. 

Seven emission lines ($H_{\alpha}(6565 A^{\circ})$, $H_{\beta}(4863
A^{\circ})$, $ OI(6302 A^{\circ})$, $OII (3727 A^{\circ})$, $ OIII
(5008 A^{\circ})$, $ NII (6585 A^{\circ})$, $SII (6718 A^{\circ}))$
are required to classify a galaxy as either star forming (SF) or  AGN.   
Only the  galaxies having all these seven emission lines with 
$I_\lambda/\sigma_{I_{\lambda}}>2$ for all the lines are considered
for classification as SF or AGN, galaxies which fail to meet this
criteria are referred to as Other galaxies. 
Here $I_\lambda$ is the emission line flux and $\sigma_{I_{\lambda}}$ is it's 
uncertainty.  We have further classified the Other galaxies as either
red   ($u-r>2.22$) or blue using the criteria   proposed  by
\citet{strat}. This  color selection
criteria  ensures that the red galaxies are `ellipticals' with 
$90 \%$ completeness.  
Most of the galaxies ($\sim 80 \%$) classified as Other
are  found to be red  galaxies.

Both the SF galaxies and AGNs show strong $H_{\alpha}$ emission. To
differentiate between them we use the BPT diagram (\citealt{baldwin})
where the logarithm of the ratio's of $[OIII]/H_\beta$ and
$[NII]/H_{\alpha}$ are plotted (Figure \ref{fig:1}). We used together
the demarcation curve provided by \citet{kauff},
$\log(\frac{OIII}{H_{\beta}})>
0.61/(\log(\frac{NII}{H_{\alpha}})-0.05)+1.3$ and the theoretical
separation curve
$\log(\frac{OIII}{H_{\beta}})>0.61/(\log(\frac{NII}{H_{\alpha}})-0.47)+1.19$
provided by \citet{kewley}. Galaxies which lie below both these two
curves are classified as SF whereas those which lie above both the
curves are classified as AGNs. Galaxies living in the intermediate
region of these two curves are discarded from further analysis.

 Table I. gives a detailed break-up of the composition of the seven
 slices.   When comparing  the filamentarity of two  different classes
 of  galaxies it is  necessary that the galaxy number density be the same
 for both the classes (Paper A).  We have ensured this by culling the
class of  galaxies which have a larger number density.

The  $H_{\alpha}$ line  was used to determine the star formation rate
(SFR) of the SF galaxies  in units of $ (M_{\odot} yr^{-1})$ using 
the relation given by \citet{hop}
$$SFR_{H\alpha}(M_{\odot} yr^{-1})=4 \pi D_{l}^2 S_{H\alpha}
\frac{10^{-0.4(r_{petro}-r_{fiber})}}{1.27 \times 10^{34}}
(\frac{S_{H\alpha}/S_{H\beta}}{2.86})^{2.114}$$
where $D_l$ is the luminosity distance, $S_{H\alpha}$ and $S_{H\beta}$
are the steller absorption corrected $H_{\alpha}$ and $H_{\beta}$
fluxes respectively and $r_{petro}$ and $r_{fiber}$ are r-band
Petrosian and fiber magnitudes respectively. The reader is referred to 
\citet{kenni} for various other definitions of the SFR. The SF
galaxies are further classified as either  high SFR or low SFR  galaxies
using the criteria $SFR \, > \, 2.7 \, M_{\odot} yr^{-1}$ which has
been chosen so that the number density of high and low SFR galaxies
are nearly the same. 

\begin{table*}{}
\caption{The definition and composition of the seven strips analyzed here.}
\vspace*{.2cm}
\begin{tabular}{|c|c|c|c|c|c|c|c|}
\hline Strip &Lambda Range&Eta Range&Total Number of galaxies(All)&Star
forming&AGN&Other&Other Blue\\ \hline 
Strip 1&$-50\leq\lambda\leq40$&$9\leq\eta\leq11$&$2922$&$774$&$189$&$1959$
&$333$\\
Strip 2&$-50\leq\lambda\leq40$&$11\leq\eta\leq
13$&$2470$&$655$&$156$&$1659$&$327$\\ 
Strip 3&$-60\leq\lambda\leq30$&$13\leq\eta\leq15$&$2296$&$649$&$153$&$1494$&$315$\\
Strip 4&$-60\leq\lambda\leq30$&$15\leq\eta\leq17$&$2300$&$644$&$133$&$1523$&$266$\\
Strip 5&$-50\leq\lambda\leq40$&$21.5\leq\eta\leq23.5$&$2454$&$702$&$152$&$1600$&$283$\\
Strip 6&$-50\leq\lambda\leq40$&$24\leq\eta\leq26$&$2541$&$671$&$181$&$1689$&$325$\\
Strip 7&$-50\leq\lambda\leq
40$&$26\leq\eta\leq28$&$2242$&$617$&$173$&$1452$&$284$\\ \hline
\end{tabular}
\end{table*}

All the strips that we have analyzed are nearly two dimensional.
The strips were all collapsed along the thickness (the smallest
dimension) to produce 2D galaxy distributions. We
use the 2D ``Shapefinder'' statistic \citep{bharad1} to quantify the
average filamentarity of the patterns in the resulting galaxy
distribution.  A detailed discussion  is
presented in Paper A ,  and we
present  only the salient features here. The reader is referred to
\citet{sahni}  for a discussion of Shapefinders in three
dimensions.

The galaxy distribution is represented as a set of 1s on a 2-D
 rectangular grid of spacing $1 \, h^{-1} {\rm Mpc} \times 1 \,h^{-1}
 {\rm Mpc}$,  empty cells are assigned a value $0$. We identify
 connected cells with a value 1 as clusters using the
 'Friends-of-Friend' (FOF) algorithm. The filamentarity of each
 cluster is quantified using the Shapefinder ${\cal F}$ defined as
\begin{equation}
{\cal F} = \frac{(P^2 - 16 S)}{(P-4 l)^2}
\end{equation}
 where $P$ and $S$ are respectively the perimeter and the area of the
 cluster, and $l$ is the grid spacing. The  Shapefinder ${\cal F}$
 has  values $0$ and $1$ for a square and filament respectively, and
 it assumes intermediate values as  a  square is deformed to a
 filament. We use the average filamentarity  
\begin{equation} 
F_2 = {\sum_{i} {\cal S}_i^2 {\cal F}_i\over\sum_{i}{\cal S}_i^2} \,. 
\end{equation}
to asses the overall filamentarity of  the clusters in the galaxy
distribution. 

The distribution of 1s corresponding to the galaxies is sparse. Only
$\sim 1 \%$ of the cells contain galaxies and there are very few
filled cells which are interconnected.  As a consequence FOF fails to
identify the large coherent structures which corresponds to filaments
in the galaxy distribution .  We overcome this by successively
coarse-graining the galaxy distribution. In each iteration of
coarse-graining all the empty cells adjacent to a filled cell are
assigned a value $1$. This causes clusters to grow, first because of
the growth of individual filled cells, and then by the merger of
adjacent clusters as they overlap. Coherent structures extending
across progressively larger length-scales are identified in
consecutive iterations of coarse-graining. Finally a transition from
many individual structures to an interconnected network is found to
occur at a filling factor $0.5-0.6$ \citep{pandey1}.

So as not to restrict our analysis to an arbitrarily
chosen level of coarse-graining, we study the average filamentarity
after each iteration of coarse-graining. The filling factor $FF$
quantifies the fraction of cells that are filled and its value
increases from $\sim 0.01$ and approaches $1$ as the coarse-graining
proceeds. We study the average filamentarity $F_2$ as a function of
the filling factor $FF$ (Figure \ref{fig:3}) as a quantitative measure
of the filamentarity at different levels of coarse-graining. The
values of $FF$ corresponding to a particular level of coarse-graining
shows a slight variation from strip to strip.  In order to combine and
compare the results from different strips, for each strip we
have interpolated $F_2$ to $7$ values of $FF$ at an uniform spacing of
$0.1$ over the interval $0.05$ to $0.65$. Coarse-graining
beyond $FF \sim 0.65$ washes away the filaments and hence we do not
include this range for our analysis.

Our method of analysis  is  similar to the Friend-of-Friends method
with varying linking length and also the density field (DF) method with a
fixed smoothing length and varying threshold density \citep{einas10}.   
All  these methods identify density enhancements in the cosmic web.
With increasing lengthscale we progress from   clusters $(\sim 1 -10 \,
{\rm Mpc})$ to  superclusters $(\sim 10 -100 \,{\rm Mpc})$
 and finally an infinite interconnected network, the cosmic web.
Often individual one dimensional  structures are identified as
filaments using a variety of criteria, for example a chain of galaxies
connecting two adjacent clusters \citep{pimb,stoi}. Instead of
focusing on the properties of individual structures, we 
quantify the overall filamentarity of the entire galaxy
distribution.  This is done over a range of filling factors. 
Figure~9 of Paper A shows that the average
length of the structures identified by our method 
is $\sim 10 \, {\rm Mpc}$  at $FF\le 0.1$ 
corresponding to galaxy clusters, $\le 100 \, {\rm Mpc}$  for  $FF \le
0.4$   corresponding to superclusters and the average length is $> 100
\, {\rm Mpc}$ for  $0.5 \le FF \le 0.6$ where we have the percolation
transition. 
We note that the percolation threshold is $FF= 0.12$  in the DF method
 \citep{einas3,einas5,einas10}  indicating that the structures  
identified by our method  are inherently thicker enclosing larger empty 
regions relative to the DF method.  Despite this, it is justified to
refer to these structures as filaments because the average
filamentarity $F_2$ of these structures is quite high
(Figure~\ref{fig:3}) for  a large range of the filling factor $FF$.    

\section*{Results and Conclusions}

\begin{figure}
\rotatebox{-90}{\scalebox{.42}{\includegraphics{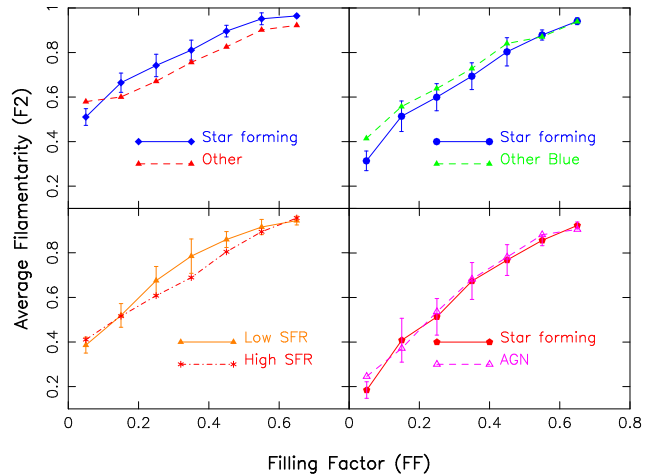}}}
\caption{Average Filamentarity as a function of Filling Factor
  comparing the different classes of galaxies  indicated in the
  panels. } 
\label{fig:3}
\end{figure}

We first compare the average filamentarity of the star forming
galaxies with that of the Other galaxies. For both classes of galaxies
we use the results from the seven different strips to compute the mean
and variance of the average filamentarity $F_2$ at uniformly chosen
values of the filling factor $FF$. The results are shown in the top
left panel of Figure \ref{fig:3}.  We find that the SF galaxies have a
higher $F_2$ compared to the Other galaxies at all values of $FF$
except at the smallest values $FF<0.1$ where the Other galaxies have a
larger $F_2$.

At each value of $FF$ we use the Student's t-test to determine
whether the differences in  $F_2$ between the SF and Other 
galaxies is
statistically significant or not. For each class of galaxies we have
seven strips which we use to calculate the mean and variance of
$F_2$. We use the Student's t-test to test the null hypothesis that
the mean $F_2$ is the same for the two classes of galaxies. 
The variance in $F_2$ is similar for the   SF and the Other galaxies,
and we use  
$s_D=\sqrt{\frac{\sum_{i\in A} (x_i-\bar x_A)^2 +\sum_{i\in B}
     (x_i-\bar x_B)^2}{N_A+N_B-2}(\frac{1}{N_A}+\frac{1}{N_B})}$ 
to  estimate the standard error for the difference in the means.
Here A and B refer to the SF and Other galaxies respectively,  
$\bar x_A$ and $\bar x_B$ refer to the mean $F_2$ for the respective
class of galaxies and  $N_A=N_B=7$ respectively refer to the number
of data points (different slices) for each class of galaxies. 
 We use $t=\frac{\bar x_A-\bar x_B}{s_D}$ to estimate the 
significance of the differences in the means. This is expected to
follow a Student's t-distribution with $12$ degrees of freedom. 
We accept the difference in the means as being statistically significant
if the probability of its occurring by chance is less than $5 \%$, 
Here we find that the difference in the mean average filamentarity
between the SF  and Other galaxies is  statistically significant at
all values of 
$FF$. 

In an earlier work \citep{pandey1} we have studied the colour and
morphology dependence of $F_2$.  We find that the blue galaxies have a
larger value of $F_2$ at nearly the entire range of $FF$ except at the
smallest values $(FF \leq 0.2$) where the red galaxies have a larger
$F_2$. A similar behaviour was also seen when the spirals were
compared with the elliptical galaxies. It was found that at large
$FF$ the spirals have a larger $F_2$ as compared to ellipticals, and
the behaviour is reversed at ($FF \le 0.25$).  Most of the SF galaxies
are blue while the Other galaxies are a mixture of blue (spiral) and
red (elliptical) galaxies.  As noted earlier, the Other galaxies are
predominantly red ellipticals ($\sim 80 \%$).  To test if the observed
difference in the filamentarity between the SF and Other galaxies is a
consequence of the fact that the SF galaxies are predominantly blue
galaxies whereas the Other galaxies are predominantly red, we have
separately compared the SF galaxies with blue galaxies drawn from the 
Other galaxies. The results are shown in the top right panel of Figure
\ref{fig:3}.  We find that the blue galaxies drawn from the 
Other galaxies have a higher $F_2$ than the SF galaxies for the
entire $FF$ range, though the difference is statistically significant
at only the smallest $FF$ value.  The error-bars for the
comparison of SF and blue Other galaxies are relatively larger than
those for the comparison of SF and the entire sample of Other
galaxies. The large error-bars do not permit us to rule out the
possibility that the excess filamentarity of the SF galaxies relative
to the Other galaxies is  a reflection of the differences in the
filamentarity of red and blue galaxies.

We next compare the  filamentarity of SF galaxies with that of  AGNs. 
The results are shown in the bottom right panel of Figure
 \ref{fig:3}. We find no statistically significant difference in the
 filamentarity of SF galaxies and AGNs except at the lowest value 
 $FF=0.05$ where the  AGNs have a larger filamentarity compared to the
 SF galaxies.

\begin{figure}
\rotatebox{-90}{\scalebox{.34}{\includegraphics{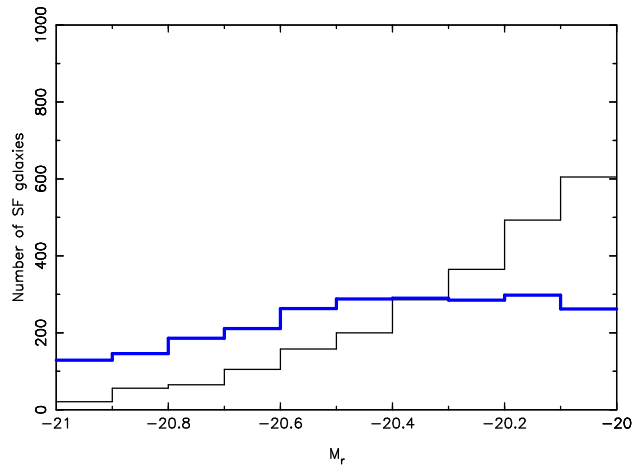}}}
\caption{Absolute magnitude distribution of high SFR (thick blue) and
  low SFR (thin black) galaxies.}
\label{fig:4}
\end{figure}

Finally we  divide the SF galaxies into two classes with equal number
of galaxies based on their star
 formation rates (SFR) and compare the filamentarity of the galaxies
 with  low SFR $(< 2.7 \, M_{\odot} {\rm yr^{-1}})$ with those with
 high SFR.  
 The results are shown in the bottom left panel of Figure \ref{fig:3}. 
 We find that there is a statistically significant difference over the
 range $0.25 \le FF \le 0.45$ corresponding to superclusters. The low
 SFR galaxies show a higher 
 filamentarity compared to  the high SFR galaxies.  
 The filamentarity of high and low
 SFR galaxies do not show any statistically significant differences at
 low and high $FF$.  

 Earlier works \citep{pandey,pandey1} show that  the low
 luminosity  galaxies have  a higher 
 filamentarity compared to the high luminosity  galaxies.  
(Figure~7, Paper~A). This is  very similar to the drop in the
 filamentarity of high SFR galaxies  compared to the low SFR ones. 
A possible explanation for the  difference in filamentarity
 observed between high SFR and low SFR   galaxies is to 
assume that   the SFR is related to luminosity, with the more luminous
 galaxies  having a higher SFR. 
To test if there actually is a relation
 between the SFR and the luminosity, we have separately considered the
 luminosity (absolute magnitude) distribution of the galaxies in the 
low SFR and high SFR classes (Figure \ref{fig:4}). We find that there is 
 a relatively  larger  number of low luminosity galaxies in the 
the low SFR class as compared to the high SFR class. Similarly there
 is a larger number of high luminosity galaxies in the high SFR class
 in comparison with the low SFR class. 

 \citet{brinch} have used a large sample of star forming galaxies
 drawn from the SDSS to study  how the SFR depends on various physical 
parameters of the individual galaxies. In  Figure~17 of their paper
 they show  evidence for a  strong correlation between the  stellar
 mass and the  SFR.  Over a large range of stellar masses $6 <
 \log(M_{\star}/M_{\odot})  <  10$  they find that 
the mean SFR increases with $M_{\star}$. The correlation breaks down
 at $ \log(M_{\star}/M_{\odot}) \ga 10$. Further, they find
 that the strong correlation between $M_{\star}$ and the  SFR is a
 recurring theme throughout their analysis.

Environmental effects are another important factor which could be
responsible for the high SFR galaxies having a lower  filamentarity.
Our earlier works using N-body simulations \citep{bharad3,pandey2}
show that the filamentarity depends on the bias. 
The filamentarity falls if the galaxies have a high bias {\it ie} the
galaxies preferentially inhabit high density regions. 
It is thus possible to explain our findings if we assume that the high
SFR galaxies preferentially inhabit denser regions compared to the low
SFR ones. 

In summary we have identified two possible explanations for the fact
that the high SFR galaxies have a less filamentary distribution than
the low SFR ones: (1.) A correlation between the SFR and individual
galaxy properties like luminosity (2.) A relation between the SFR and
environmental effects like the density.  Possibly both of these are
interconnected and are simultaneously at play.  It is interesting to
discuss these two possibilities in the light of other related
findings.  It is well known that the more luminous galaxies
preferentially inhabit denser regions (e.g. \citealt{einas8},
\citealt{einas2}, \citealt{einas5}, \citealt{park2},) and exhibit
stronger clustering strength compared to the fainter ones
(e.g. \citealt{hamil1}, \citealt{davis1}, \citealt{white},
\citealt{park}, \citealt{love}, \citealt{guzo}, \citealt{nor},
\citealt{zehavi}). This effect is quantified through a luminosity bias
relation \citep{beno, nor, teg2, pandey2}.  Earlier studies of the
environment dependence \citep{lewis,gomez,balogh} all show a
suppression of star formation activity in high density regions. While
these works all find a decrease in the fraction of SF galaxies within
a distance of $2-3$ virial radius from the center of clusters, the
effect of density on the SFR distribution is not clear. \citet{gomez}
find that the SFR distribution is strongly shifted to lower values in
high density environments, the effect being most pronounced for the
strongest star forming galaxies whereas \citet{balogh} find that the
SFR distribution is independent of environment. \citet{einas6} find
that the fraction of SF galaxies in rich groups/clusters in high
density regions of rich supercluster is smaller than what is found in
poor superclusters in the field. In their analysis superclusters are
connected non-percolating systems with densities above a certain
density threshold. \citet{einas7} also find that the more luminous and
richer superclusters have a higher degree of filamentarity compared to
the poor ones. \citet{einas9} and \citet{einas6} show that both the
local and the global environment are possibly important in influencing
morphological properties and star formation. Recently \citet{lee} have
studied the correlation between the large scale environmnet of
galaxies and their physical properties in the SDSS and the 2Mass
Redshift Survey. They find that the physical parameters related to the
recent star formation history are linked to the shear of the large
scale environment of galaxies. A study of star formation along the
Pisces-Cetus Supercluster filaments \citep{porter} finds that though
the SFR and the fraction of SF galaxies declines steadily towards the
cores of clusters, there is an increased SF activity in a narrow
distance range around $1.5$ to $2$ times the virial radius of the
cluster involved. A study of the environment and clustering properties
of star-burst galaxies in the 2dFGRS \citep{owers} finds that a
significant fraction of star-burst galaxies show morphological
evidence for ongoing or recent tidal interaction or merger.

There appears to be no consensus on how star formation depends on the
environment and some of the findings appears to be at odds with our
results, we note that the associated length-scales are quite
different. While the earlier studies probed groups and clusters of
galaxies {\it ie.}  length-scales less than a few Mpc, the structures
identified by our method have lengths of the order of $100 \, {\rm
Mpc/h}$ and thickness of the order $5 \, {\rm Mpc/h}$ (Figure 9, Paper A)
for the range of $FF$ where the difference in filamentarity is
statistically significant. Our interpretation is consistent with the
recent analysis by \citet{sio} who has shown that the high SF galaxies
at $z \approx0.24$ are more strongly clustered than the low SF
galaxies.

\section{Acknowledgment}

BP acknowledges Raghunathan Srianand and Swara Ravindranath for useful
discussions. The SDSS DR5 data was downloaded from the SDSS skyserver
http://cas.sdss.org/dr5/en/

    Funding for the creation and distribution of the SDSS Archive has been 
provided by the Alfred P. Sloan Foundation, the Participating 
Institutions, the National Aeronautics and Space Administration, the 
National Science Foundation, the U.S. Department of Energy, the Japanese 
Monbukagakusho, and the Max Planck Society. The SDSS Web site is 
http://www.sdss.org/.

    The SDSS is managed by the Astrophysical Research Consortium (ARC) for 
the Participating Institutions. The Participating Institutions are The 
University of Chicago, Fermilab, the Institute for Advanced Study, the 
Japan Participation Group, The Johns Hopkins University, the Korean 
Scientist Group, Los Alamos National Laboratory, the Max-Planck-Institute 
for Astronomy (MPIA), the Max-Planck-Institute for Astrophysics (MPA), New 
Mexico State University, University of Pittsburgh, Princeton University, 
the United States Naval Observatory, and the University of Washington.

\end{document}